\documentclass{article}
\usepackage[intlimits]{amsmath}
\usepackage[cp1251]{inputenc}
\usepackage[T2A]{fontenc}
\usepackage[russian,english]{babel}
\usepackage[dvips]{color}
\usepackage{graphics}
\begin{document}
\center{\Large \bf{Generation of polarization squeezed light in
periodically poled nonlinear crystal}}

{\center V. G. Dmitriev}\\

{\center R\&D Institute "Polyus"\\ Vvedenskogo st., 3, Moscow
117342, Russia}

{\center R. Singh}\\

{\center General Physics Institute of RAS\\ Vavilov st., 38,
Moscow 119991, Russia\\ E-mail: ranjit.singh@mail.ru}

\abstract{Theoretical analysis is presented on quantum state
evolution of polarization light waves at frequencies $\omega_{o}$
and $\omega_{e}$ in a periodically poled nonlinear crystal (PPNC).
It is shown that the variances of all the four Stokes parameters
can be squeezed.}
\bigskip


\section{Introduction}
During the last decade much attention has been paid to the
realization of quantum information protocols [1] such as quantum
teleportation, quantum cryptography. These protocols are based on
the methods of nonlinear quantum optics. Nonlinear optical sources
[2] play an important role in the generation of nonclassical
states of light [3] and realization of optical quantum information
protocols. The nonlinear optical processes such as degenerate
parametric down conversion (type I and II processes) [2,3] and the
Kerr effect [2,3] are used to create nonclassical states of light
(squeezed states, polarization squeezed states and entangled
states). The variance of one of the four Stokes parameters of
polarization squeezed light is less than the corresponding value
for the coherent state. Traditionally the degenerate parametric
process (type II) and the Kerr effect are responsible for the
generation of polarization squeezed states or polarization
entangled states in ordinary nonlinear crystals. One can achieve
suppression of variances of at least one of the Stokes parameters
[5-9] $(\hat{S}_{0},\hat{S}_{1},\hat{S}_{2},\hat{S}_{3})$ in the
type II process by using ordinary nonlinear crystal. Most of the
quantum information protocols are based on type II process [1] and
Kerr effect, which are used to generate entangled states. The
entangled states are used in the realization of quantum
teleporation and quantum cryptography protocols. Experiments on
quantum teleportation and quantum cryptography are performed by
using ordinary nonlinear optical crystals with second and third
order nonlinear susceptibilities. In the past few years, some
experiments in the creation and realization of nonclassical and
entangled states are performed by using PPNCs with second order
nonlinear susceptibilities. The PPNCs [10,11], which have many
interesting advantages as compared to ordinary nonlinear crystals
were proposed by Bloembergen and co-authors in 1962. The main
advantages of PPNCs against ordinary nonlinear crystals are: the
quasi-phase-matching condition between the interacting waves; the
highest nonlinear susceptibility coefficient can be used;
multi-mode interaction of optical waves.

Recent experiments on quantum noise reduction [12,13] and the
generation of entangled states [14-16] promising the applications
of PPNCs in the realization of optical quantum information
protocols. These experiments were based on type I [12-14] and II
[16] processes. It should be noted that the parametric down
conversion (type I) and frequency sum generation processes have
been studied theoretically [see for instance, [4] and the
references therein] and experimentally [12-16] very well. The
experiment on generation of time-bin and energy-bin entangled
states using the parametric down conversion process (type I, i.e.
$2\omega_{e}=\omega_{e}+\omega_{e}$) in a PPNC was demonstrated by
Gisin and co-workers [14]. In this experiment authors claimed the
higher energy conservation from the fundamental beam $2\omega_{e}$
to modes with frequencies $\omega_{e}$ and $\omega_{e}$. In a
PPNC, the possibilities of type II process is much more
complicated as compared to ordinary nonlinear crystals. The
following are the some of possible type II nonlinear processes
which can be realized in a PPNC [4]: (a)
$\omega_{o}+\omega_{e}=2\omega_{e}$ (one can achieve suppression
of maximum three variances of Stokes parameters under certain
conditions. The same can be achieved in an ordinary nonlinear
crystal), (b) $\omega_{o}+\omega_{e}=2\omega_{e}$ and
$2\omega_{e}+\omega_{o}=3\omega_{e}$ (one can achieve suppression
of all the four variances of Stokes parameters under certain
conditions. The same can not be achieved in an ordinary nonlinear
crystal), (c) $\omega_{o}+\omega_{e}=2\omega_{e}$,
$2\omega_{e}+\omega_{o}=3\omega_{e}$,
$3\omega_{e}+\omega_{e}=4\omega_{o}$ and
$2\omega_{e}+2\omega_{e}=4\omega_{o}$ (one can achieve higher
suppression of all the four variances of Stokes parameters under
certain conditions as compared to case (b). The same can not be
achieved in an ordinary nonlinear crystal). The type II process
(a) in PPNC can be accompanied by considering other nonlinear
processes ((b) or (c)) [4]. The nonlinear process (a) differs with
the same nonlinear process in an ordinary bulk crystal by a higher
energy conversion rate of fundamental mode ($2\omega_{e}$) into
degenerate (orthogonal) modes ($\omega_{o}$ and $\omega_{e}$)
[16]. The recent experiment on generation of polarization
entangled states (type II process (a)) [16] demonstrated a high
energy conversion rate from fundamental mode ($2\omega_{e}$) into
degenerate (orthogonal) modes ($\omega_{o}$ and $\omega_{e}$). The
nonlinear processes (b) or (c) can be realized in a single PPNC
but not in a single ordinary nonlinear crystal. All these
nonlinear processes (b) or (c) can be quasi-phase-matched at
certain coherent lengths [4]. Here we will study the generation of
polarization squeezed states based on type II process (c) in PPNC
with second order nonlinear susceptibility.

The main goal of this work is to show that PPNCs can suppress all
the four variances of Stokes parameters below the standard quantum
limit. The theoretical work that we present here is the first, to
the best of our knowledge, to propose PPNCs for the generation of
polarization squeezed light.

The structure of the paper is as follows. Section 2 describes the
optical nonlinear processes and their Heisenberg equations of
motions. Section 3 studies the behaviour of mean photon numbers of
degenerate polarization (orthogonal) modes at frequencies
$\omega_{o}$ and $\omega_{e}$. Section 4 analyzes the variances of
Stokes parameters. The final section summarizes the results
obtained in sections 3 and 4.

\section{Equations of motions}
We consider the five-frequency interaction of co-propagating light
waves in a PPNC (see Fig.1). The four interaction processes of
light waves at frequencies $\omega_{o}$, $\omega_{e}$,
$2\omega_{e}$, $3\omega_{e}$, and $4\omega_{o}$ are [4]
\begin{eqnarray}
\omega_{o}+\omega_{e}=2\omega_{e},\nonumber \\
\delta{k}_{1}=k_{2e}-k_{1o}-k_{1e}+m_{1}G_{1}=\Delta{k}_{1}+m_{1}G_{1},\\
\omega_{o}+2\omega_{e}=3\omega_{e}, \nonumber \\
\delta{k}_{2}=k_{3e}-k_{1o}-k_{2e}+m_{2}G_{2}=\Delta{k}_{2}+m_{2}G_{2},\\
\omega_{e}+3\omega_{e}=4\omega_{o}, \nonumber \\
\delta{k}_{3}=k_{4e}-k_{1e}-k_{3e}+m_{3}G_{3}=\Delta{k}_{3}+m_{3}G_{3},\\
2\omega_{e}+2\omega_{e}=4\omega_{o}, \nonumber \\
\delta{k}_{4}=k_{4e}-2k_{2e}+m_{4}G_{4}=\Delta{k}_{4}+m_{4}G_{4},
\end{eqnarray}
where $\Delta{k}_{j=1,2,3,4}$ is a phase mismatch for an ordinary
nonlinear bulk crystal; $k_{jn}$ is a wave number of interacting
modes ($n=o$ stands for ordinary and $n=e$ for extraordinary);
$G_{j}=2\pi \Lambda_{j}^{-1}$ is the modulus of the vector of a
reciprocal lattice with period $\Lambda_{j}$; $m_{j}=\pm 1, \pm
3,\ldots $, is the quasi-phase-matched order.

The nonlinear process (1) describes the splitting of a photon of
frequency $2\omega_{e}$ into two photons of orthogonal
polarizations with degenerate frequencies $\omega_{o}$ and
$\omega_{e}$. The second process (2) describes the sum frequency
generation, i.e. photon of frequency $2\omega_{e}$ combines with
the photon of frequency $\omega_{o}$, which gives rise to a photon
of frequency $3\omega_{e}$. The third (3) and fourth (4) processes
are responsible for the generation of the fourth harmonic with two
different ways, i.e. a photon of frequency $\omega_{e}$ and a
photon of frequency $3\omega_{e}$ combine, and a photon with
frequency $4\omega_{o}$ appears as a result or the same can be
achieved by the combination of two photons with frequencies
$2\omega_{e}$.

It has been shown that the nonlinear processes (1-4) can be
simultaneously quasi-phase-matched [4] in a single domain
structure $(G_{1}=G_{2}=G_{3}=G_{4})$ or at certain coherent
lengths $L_{coh}^{j}$. The processes (1-4) under study can be
simultaneously quasi-phase-matched with the following relationship
\begin{eqnarray}
L_{coh}^{1,2}=9L_{coh}^{3,4}.
\end{eqnarray}
The processes (1-4) can be described by the following interaction
Hamiltonian
\begin{eqnarray}
\hat{H}_{I}(z)=\hbar
g(z)[\xi_{1}\hat{a}_{1o}\hat{a}_{1e}\hat{a}_{2e}^{+}e^{i\Delta
k_{1}z}+\xi_{2}\hat{a}_{1o}\hat{a}_{2e}\hat{a}_{3e}^{+}e^{i\Delta
k_{2}z}+\xi_{3}\hat{a}_{1e}\hat{a}_{3e}\hat{a}_{4o}^{+}e^{i\Delta
k_{3}z}+\xi_{4}\hat{a}_{2e}^{2}\hat{a}_{4o}^{+}e^{i\Delta
k_{4}z}+HC],
\end{eqnarray}
where $\hbar$ is Planck's constant, $\hat{a}_{jn}
(\hat{a}_{jn}^{+})$ is the annihilation (creation) operator of
photon of $jn$th mode at frequency $j\omega_{n}$; $\xi_{j}$ is the
nonlinear coupling coefficient; $g(z)$ is the periodic function
equal to $+1$ or $-1$ at the domain thickness $l=\Lambda /2$; HC
denotes Hermitian conjugate. The operators $\hat{a}_{jn}
(\hat{a}_{jn}^{+})$ obey the following commutation rules
\begin{eqnarray}
[\hat{a}_{jn},\hat{a}_{pn}^{+}]=\delta_{jn,pn},&
\mathrm{with}&{j,p=1,2,3,4}
\end{eqnarray}
The interaction Hamiltonian (6) can be averaged over the period
$\Lambda$, if the interaction length $z$ is much more than the
period of modulation $\Lambda$, i.e. $z\gg \Lambda$. Then the
interaction Hamiltonian (6) takes the form
\begin{eqnarray}
\hat{H}_{I}=\hbar
[\gamma_{1}\hat{a}_{1o}\hat{a}_{1e}\hat{a}_{2e}^{+}+\gamma_{2}\hat{a}_{1o}\hat{a}_{2e}\hat{a}_{3e}^{+}+\gamma_{3}\hat{a}_{1e}\hat{a}_{3e}\hat{a}_{4o}^{+}+\gamma_{4}\hat{a}_{2e}^{2}\hat{a}_{4o}^{+}+HC],
\end{eqnarray}
where
$$\gamma_{j}=\frac{\xi_{j}}{\Lambda}\int_{-\Lambda/2}^{+\Lambda/2}g(z)\exp{(\pm
i\Delta k_{j}z )}=2\xi_{j}/(\pi m_{j}).$$ The Heisenberg operator
equations corresponding to the interaction Hamiltonian (8) are
given by
\begin{eqnarray}
i\frac{d\hat{a}_{1o}}{dz}=\gamma_{2}\hat{a}_{2e}^{+}\hat{a}_{3e}+\gamma_{1}\hat{a}_{1e}^{+}\hat{a}_{2e},\nonumber
\\
i\frac{d\hat{a}_{1e}}{dz}=\gamma_{3}\hat{a}_{3e}^{+}\hat{a}_{4o}+\gamma_{1}\hat{a}_{1o}^{+}\hat{a}_{2e},\nonumber
\\
i\frac{d\hat{a}_{2e}}{dz}=2\gamma_{4}\hat{a}_{4o}\hat{a}_{2e}^{+}+\gamma_{2}\hat{a}_{1o}^{+}\hat{a}_{3e}+\gamma_{1}\hat{a}_{1e}\hat{a}_{1o},\nonumber
\\
i\frac{d\hat{a}_{3e}}{dz}=\gamma_{2}\hat{a}_{1o}\hat{a}_{2e}+\gamma_{3}\hat{a}_{1e}^{+}\hat{a}_{4o},\nonumber
\\
i\frac{d\hat{a}_{4o}}{dz}=\gamma_{4}\hat{a}_{2e}^{2}+\gamma_{3}\hat{a}_{1e}\hat{a}_{3e}.
\end{eqnarray}
Assuming the pump modes at frequencies $\omega_{2e}$,
$\omega_{4o}$ are classical and non-depleted at the input of a
PPNC, i.e.
\begin{eqnarray}
\hat{a}_{2e}=A_{2e},\\ \hat{a}_{4o}=A_{4o},
\end{eqnarray}
where $A_{2e,4o}=e^{i\pi/2}$ are the complex amplitudes of the
pump modes, we obtain the following linear system of equations,
after the substitution of quantities (10) and (11) into (9):
\begin{eqnarray}
\frac{d\hat{a}_{1o}}{dz}=-\gamma_{2}\hat{a}_{3e}+\gamma_{1}\hat{a}_{1e}^{+},\nonumber
\\
\frac{d\hat{a}_{1e}}{dz}=\gamma_{3}\hat{a}_{3e}^{+}+\gamma_{1}\hat{a}_{1o}^{+},\nonumber
\\
\frac{d\hat{a}_{3e}}{dz}=\gamma_{2}\hat{a}_{1o}+\gamma_{3}\hat{a}_{1e}^{+}.
\end{eqnarray}
For simplicity, we introduce a normalized interaction length
parameter $\zeta$
\begin{eqnarray}
\zeta=z\gamma_{1}.
\end{eqnarray}
The quantity (13) is introduced into the set of equations (12),
which reduce after straightforward algebra to the set of equations
\begin{eqnarray}
\frac{d\hat{a}_{1o}}{d\zeta}=-k_{2}\hat{a}_{3e}+\hat{a}_{1e}^{+},\nonumber
\\
\frac{d\hat{a}_{1e}}{d\zeta}=k_{1}\hat{a}_{3e}^{+}+\hat{a}_{1o}^{+},\nonumber
\\
\frac{d\hat{a}_{3e}}{d\zeta}=k_{2}\hat{a}_{1o}+k_{1}\hat{a}_{1e}^{+}.
\end{eqnarray}
where $k_{1}=\gamma_{3}/\gamma_{1}$ and
$k_{2}=\gamma_{2}/\gamma_{1}$. The set of linear equations (14) is
solved by applying the Laplace transformation:
\begin{eqnarray}
\hat{a}_{1o}(\zeta)=\lambda_{11}(\zeta)\hat{a}_{1o}+\lambda_{12}(\zeta)\hat{a}_{1e}^{+}+\lambda_{13}(\zeta)\hat{a}_{3e},\nonumber
\\
\hat{a}_{1e}(\zeta)=\lambda_{21}(\zeta)\hat{a}_{1o}^{+}+\lambda_{22}(\zeta)\hat{a}_{1e}+\lambda_{23}(\zeta)\hat{a}_{3e}^{+},\nonumber
\\
\hat{a}_{3e}(\zeta)=\lambda_{31}(\zeta)\hat{a}_{1o}+\lambda_{32}(\zeta)\hat{a}_{1e}^{+}+\lambda_{33}(\zeta)\hat{a}_{3e}.
\end{eqnarray}
where $\hat{a}_{jn}=\hat{a}_{jn}(0)$ and
\begin{eqnarray}
q=\sqrt{1+k_{1}^{2}-k_{2}^{2}},\nonumber \\
\lambda_{11}(\zeta)=\frac{1}{q^{2}}(-k_{1}^{2}\cosh{q\zeta}+q^{2}\cosh{q\zeta}+k_{1}^{2}),\nonumber
\\
\lambda_{12}(\zeta)=\frac{1}{q^{2}}(-k_1k_2\cosh{q\zeta}+k_1k_2+q\sinh{q\zeta}),\nonumber
\\
\lambda_{13}(\zeta)=\frac{1}{q^{2}}(-qk_2\sinh{q\zeta}+k_1\cosh{q\zeta}-k_1),\nonumber
\\
\lambda_{21}(\zeta)=\frac{1}{q^{2}}(q\sinh{q\zeta}-k_{1}k_{2}+k_{1}k_{2}\cosh{q\zeta}),\nonumber
\\
\lambda_{22}(\zeta)=\frac{1}{q^{2}}(k_{2}^{2}\cosh{q\zeta}+q^{2}\cosh{q\zeta}-k_{2}^{2}),\nonumber
\\
\lambda_{23}(\zeta)=\frac{1}{q^{2}}(qk_{1}\sinh{q\zeta}-k_{2}\cosh{q\zeta}+k_2),\nonumber
\\
\lambda_{31}(\zeta)=\frac{1}{q^{2}}(qk_{2}\sinh{q\zeta}-k_{1}+k_{1}\cosh{q\zeta}),\nonumber
\\
\lambda_{32}(\zeta)=\frac{1}{q^{2}}(qk_{1}\sinh{q\zeta}+k_{2}\cosh{q\zeta}-k_{2}),\nonumber
\\
\lambda_{33}(\zeta)=\frac{1}{q^{2}}(1+q^{2}\cosh{q\zeta}-\cosh{q\zeta}).
\end{eqnarray}
If $k1=k2=0$, then the solution (15) corresponds to the
conventional degenerate parametric down conversion process (type
II) [3]. Using (15) we can calculate the statistical properties of
interacting modes with frequencies $\omega_{o}$, $\omega_{e}$,
$3\omega_{e}$ at normalized interaction length $\zeta$.
\section{Evolution of mean photon numbers}
The evolution of mean photon number of degenerate polarization
modes at frequencies $\omega_{o}$ and $\omega_{e}$ in a PPNC are
calculated by using (15) for the following initial conditions at
the input of the PPNC: the two orthogonal modes are in polarized
coherent states $|\alpha_{1o}>$, $|\alpha_{1e}>$,
$\alpha_{1o,e}=|\alpha_{1o,e}|e^{i\phi_{1o,e}}$, where
$|\alpha_{1o,1e}|=\sqrt{<N_{1o,e}(0)>}$ and the wave at frequency
$3\omega_{e}$ is in the vacuum state $|0>$
\begin{eqnarray}
<\hat{N}_{1o,e}(\zeta)>=<\alpha_{1o}|<\alpha_{1e}|<0|\hat{a}_{1o,e}^{+}(\zeta)\hat{a}_{1o,e}(\zeta)|0>|\alpha_{1e}>|\alpha_{1o}>
\end{eqnarray}
The expressions for mean photons (17) are
\begin{eqnarray}
<N_{1o}(\zeta)>=\lambda_{12}^{2}(\zeta)+\lambda_{12}^{2}(\zeta)|\alpha_{1e}|^{2}+2\lambda_{11}(\zeta)\lambda_{12}(\zeta)\cos{(\phi_{1o}+\phi_{1e})}+\lambda_{11}^{2}(\zeta)|\alpha_{1o}|^{2},\\
<N_{1e}(\zeta)>=\lambda_{21}^{2}(\zeta)+\lambda_{23}^{2}(\zeta)+\lambda_{22}^{2}(\zeta)|\alpha_{1e}|^{2}+2\lambda_{21}(\zeta)\lambda_{22}(\zeta)\cos{(\phi_{1o}+\phi_{1e})}+\lambda_{21}^{2}(\zeta)|\alpha_{1o}|^{2}.
\end{eqnarray}
It is well known that the parametric down conversion process (type
II) depends upon the initial phases $\phi_{1o,e}$ of the polarized
(orthogonal) coherent states. The values of mean photon numbers
(18) and (19) depend upon the sum of initial phases, i.e.
$\phi_{1o}+\phi_{1e}$. Under the condition
$\phi_{1o}+\phi_{1e}=2\pi s$, the mean photon numbers (18) and
(19) start increasing rapidly with the growth of interaction
length $\zeta$, under the condition $\phi_{1o}+\phi_{1e}=\pi s$,
they start decreasing and then monotnically increasing (see Fig.
2) and $s=\pm 1,\pm 2,\ldots .$

Fig. 2 demonstrates the dependence of mean photon numbers
$<N_{1o}>$ and $<N_{1e}>$ on the normalized interaction length
under the condition $\phi_{1o}+\phi_{1e}=\pi$. It is seen that the
behaviour of mean photon numbers at frequencies $\omega_{o}$
(curve $1$) and $\omega_{e}$ (curve $2$) are quiet different from
the evolution of the mean photon number for the case of parametric
down conversion (type II) in an ordinary nonlinear crystal. The
mean photon number (curve 3) for the case of an ordinary nonlinear
crystal is calculated by putting $k_{1}=k_{2}=0$ into the
expressions (18) and (19) under the same initial conditions.

The difference between the evolutions of mean photon numbers (18)
(curve $1$) and (19) (curve $2$) illustrated in Fig. 2 is related
with the interaction process (2), which is responsible for the sum
harmonic generation, i.e, photon of frequency $\omega_{o}$
combines with the photon of pump frequency $2\omega_{e}$, which
gives rise to photon with frequency $3\omega_{e}$. This can be
easily seen by evaluating the mean photon numbers
$<N_{3e}(\zeta)>$ using (15) under the same initial conditions
\begin{eqnarray}
<N_{3e}(\zeta)>=\lambda_{32}^{2}(\zeta)+\lambda_{32}^{2}(\zeta)|\alpha_{1e}|^{2}+2\lambda_{31}(\zeta)\lambda_{32}(\zeta)\cos{(\phi_{1o}+\phi_{1e})}+\lambda_{31}^{2}(\zeta)|\alpha_{1o}|^{2}.
\end{eqnarray}
The curve (4) of Fig.2 illustrates the growth of the mean photon
number $<N_{3e}>$ as the normalized interaction length increases.

At earlier stages (normalized interaction length $\zeta\approx
0\div 0.4$), the mean photon number of frequency $\omega_{e}$
decreases (see Fig. 2, (curve 2)) and later starts increasing
monotonically. The decreasing of mean photon number $<N_{1e}>$ is
connected with the nonlinear process (3), which is responsible for
the sum harmonic generation at frequency $4\omega_{o}$ and is
realized by the combination of photons with frequencies
$\omega_{e}$ and $3\omega_{e}$. The fourth harmonic generation
process (3) is complicated as compared to (1) and (2). So, the
interaction process (3) starts acting at later stages of
interaction as compared to the interaction process (2).

The calculations of photon variances $<\Delta
N_{1o,e}^{2}(\zeta)>$ of polarization modes with frequencies
$\omega_{o}$ and $\omega_{e}$ have shown that they are
super-Poissonian [3]. The same can be seen in the ordinary
parametric down conversion process (type II) in ordinary nonlinear
crystals by putting the nonlinear coupling coefficients $k_{1}$,
$k_{2}$ equal to $0$ in the expressions (16) and substituting
$\hat{a}_{3}(0)=0$ into the solution (15).

\section{Stokes parameters}
The polarization properties of orthogonal modes at frequencies
$\omega_{o}$ and $\omega_{e}$ can be analyzed by the Stokes
parameters [8]
\begin{eqnarray}
\hat{S}_{0}(\zeta)=\hat{a}_{1o}^{+}(\zeta)\hat{a}_{1o}(\zeta)+\hat{a}_{1e}^{+}(\zeta)\hat{a}_{1e}(\zeta),
\nonumber \\
\hat{S}_{1}(\zeta)=\hat{a}_{1o}^{+}(\zeta)\hat{a}_{1o}(\zeta)-\hat{a}_{1e}^{+}(\zeta)\hat{a}_{1e}(\zeta),\nonumber
\\
\hat{S}_{2}(\zeta)=\hat{a}_{1o}^{+}(\zeta)\hat{a}_{1e}(\zeta)+\hat{a}_{1e}^{+}(\zeta)\hat{a}_{1o}(\zeta),\nonumber
\\
\hat{S}_{3}(\zeta)=i[\hat{a}_{1e}^{+}(\zeta)\hat{a}_{1o}(\zeta)-\hat{a}_{1o}^{+}(\zeta)\hat{a}_{1e}(\zeta)],
\end{eqnarray}
The Stokes parameters (21) obey the commutation relations of the
SU(2) algebra [8]
\begin{eqnarray}
[\hat{S}_{0}(\zeta),\hat{S}_{1,2,3}(\zeta)]=0;&
[\hat{S}_{1}(\zeta),\hat{S}_{2}(\zeta)]=2i\hat{S}_{3}(\zeta);
\nonumber\\
{[\hat{S}_{2}(\zeta),\hat{S}_{3}(\zeta)]}=2i\hat{S}_{1}(\zeta);&
[\hat{S}_{3}(\zeta),\hat{S}_{1}(\zeta)]=2i\hat{S}_{2}(\zeta).
\end{eqnarray}
The Heisenberg uncertainty relation for the Stokes parameters (21)
is given by [8] $$ <\bigtriangleup
\hat{S}_{i}^{2}(\zeta)><\bigtriangleup \hat{S}_{j}^{2}(\zeta)>
\geq |<\hat{S}_{k}>|^{2}, (i,j,k=1,2,3) (i\neq j\neq k). $$
\begin{eqnarray}
\hat{S}_{0,1}(\zeta)=p_{0\pm}(\zeta)+p_{1\pm}(\zeta)\hat{a}_{3e}^{+}\hat{a}_{3e}+p_{2\pm}(\zeta)\{\hat{a}_{1e}\hat{a}_{3e}+\hat{a}_{1e}^{+}\hat{a}_{3e}^{+}\}+p_{3\pm}(\zeta)\hat{a}_{1e}^{+}\hat{a}_{1e}
+p_{4\pm}(\zeta)\{
\hat{a}_{1o}\hat{a}_{3e}^{+}+\hat{a}_{1o}^{+}\hat{a}_{3e}\}\nonumber
\\+p_{5\pm}(\zeta)\{\hat{a}_{1o}\hat{a}_{1e}+\hat{a}_{1o}^{+}\hat{a}_{1e}^{+}
\}+p_{6\pm}(\zeta)\hat{a}_{1o}^{+}\hat{a}_{1o},\\
\hat{S}_{2,3}(\zeta)=i^{0,1}[q_{0}(\zeta)\{\hat{a}_{3e}^{2}\pm
\hat{a}_{3e}^{2+}\}+q_{1}(\zeta)\{ \hat{a}_{1e}^{+}\hat{a}_{3e}\pm
\hat{a}_{1e}\hat{a}_{3e}^{+}\}+q_{2}(\zeta)\{\hat{a}_{1e}^{+2}\pm
\hat{a}_{1e}^{2}\}+q_{3}(\zeta)\{\hat{a}_{1o}\hat{a}_{3e}\pm
\hat{a}_{1o}^{+}\hat{a}_{3e}^{+}\}\nonumber
\\
+q_{4}(\zeta)\{\hat{a}_{1o}\hat{a}_{1e}^{+}\pm
\hat{a}_{1o}^{+}\hat{a}_{1e}\}+q_{5}(\zeta)\{\hat{a}_{1o}^{2}\pm
\hat{a}_{1o}^{2+} \}],
\end{eqnarray}
where $i^{0}=1$, $i^{1}=i$ stand for $\hat{S}_{2}(\zeta)$,
$\hat{S}_{3}(\zeta)$ and
\begin{eqnarray}
p_{0\pm}(\zeta)=\lambda_{12}^{2}(\zeta)\pm
\lambda_{21}^{2}(\zeta)\pm \lambda_{23}^{2}(\zeta),\nonumber\\
p_{1\pm}(\zeta)=\lambda_{13}^{2}(\zeta)\pm
\lambda_{23}^{2}(\zeta),\nonumber\\
p_{2\pm}(\zeta)=\lambda_{12}(\zeta)\lambda_{13}(\zeta)\pm
\lambda_{22}(\zeta)\lambda_{23}(\zeta),\nonumber\\
p_{3\pm}(\zeta)=\lambda_{12}^{2}(\zeta)\pm
\lambda_{22}^{2}(\zeta),\nonumber\\
p_{4\pm}(\zeta)=\lambda_{11}(\zeta)\lambda_{13}(\zeta)\pm
\lambda_{21}(\zeta)\lambda_{23}(\zeta),\nonumber\\
p_{5\pm}(\zeta)=\lambda_{11}(\zeta)\lambda_{12}(\zeta)\pm
\lambda_{21}(\zeta)\lambda_{22}(\zeta),\nonumber\\
p_{6\pm}(\zeta)=\lambda_{11}^{2}(\zeta)\pm
\lambda_{21}^{2}(\zeta),\nonumber\\
q_{0}(\zeta)=\lambda_{13}(\zeta)\lambda_{23}(\zeta),\nonumber\\
q_{1}(\zeta)=\lambda_{13}(\zeta)\lambda_{22}(\zeta)+\lambda_{12}(\zeta)\lambda_{23}(\zeta),\nonumber\\
q_{2}(\zeta)=\lambda_{12}(\zeta)\lambda_{22}(\zeta),\nonumber\\
q_{3}(\zeta)=\lambda_{13}(\zeta)\lambda_{21}(\zeta)+\lambda_{11}(\zeta)\lambda_{23}(\zeta),\nonumber\\
q_{4}(\zeta)=\lambda_{12}(\zeta)\lambda_{21}(\zeta)+\lambda_{11}(\zeta)\lambda_{22}(\zeta),\nonumber\\
q_{5}(\zeta)=\lambda_{11}(\zeta)\lambda_{21}(\zeta).
\end{eqnarray}
Further for simplicity and clearness, we will write
$p_{j}(\zeta)_{(j=0,1,2,3,4,5,6)}$ as $p_{j}$ and
$q_{j}(\zeta)_{(j=0,1,2,3,4,5)}$ as $q_{j}$. The expressions for
the variances of the Stokes parameters (23) and (24) become very
lengthy. So, we write down the expressions of variances for the
Stokes parameter $\hat{S}_{j}(\zeta)$ under the same initial
conditions applied for expressions (17)-(20)
\begin{eqnarray}
<\bigtriangleup
\hat{S}_{j}^{2}(\zeta)>=<\hat{S}_{j}^{2}(\zeta)>-<\hat{S}_{j}(\zeta)>^{2},
(j=0,1,2,3)
\end{eqnarray}
The variances of Stokes parameters (26) are normalized by dividing
them by the variance of the Stoke's parameter $\hat{S}_{0}(0)$,
i.e.
\begin{eqnarray}
V_{j}(\zeta)=\frac{<\bigtriangleup
\hat{S}_{j}^{2}(\zeta)>}{<\Delta{\hat{S}_{0}}^{2}(0)>},
\end{eqnarray}
where
$<\Delta{\hat{S}_{0}}^{2}(0)>=|\alpha_{1o}|^{2}+|\alpha_{1e}|^{2}$.
For the case of coherent light, the relative variances (27) take
values $1$. If at least one of the relative variances (27) takes
value less than $1$, the light is said polarization squeezed. The
relative variances $V_{j}(\zeta)$ read
\begin{eqnarray}
V_{0,1}(\zeta)=\frac{1}{<\Delta{\hat{S}_{0}}^{2}(0)>}[p_{0\pm}^{2}+p_{2\pm}^{2}+p_{5\pm}^{2}+(p_{2\pm}^{2}+2p_{0\pm}p_{3\pm}+p_{3\pm}^{2}+p_{5\pm}^{2})|\alpha_{1e}|^{2}\nonumber
\\
+p_{3\pm}^{2}|\alpha_{1e}|^{4}+2(p_{2\pm}p_{4\pm}+2p_{0\pm}p_{5\pm}+p_{3\pm}p_{5\pm}+p_{5\pm}p_{6\pm})|\alpha_{1o}||\alpha_{1e}|\cos(\phi_{1o}+\phi_{1e})\nonumber
\\
+4p_{3\pm}p_{5\pm}|\alpha_{1o}||\alpha_{1e}|^{3}\cos{(\phi_{1o}+\phi_{1e})}+2p_{5\pm}^{2}|\alpha_{1o}|^{2}|\alpha_{1e}|^{2}\cos{2(\phi_{1o}+\phi_{1e})}\nonumber
\\
+(p_{4\pm}^{2}+p_{5\pm}^{2}+2p_{0\pm}p_{6\pm}+p_{6\pm}^{2})|\alpha_{1o}|^{2}+2(p_{5\pm}^{2}+p_{3\pm}p_{6\pm})|\alpha_{1o}|^{2}|\alpha_{1e}|^{2}\nonumber
\\
+4p_{5\pm}p_{6\pm}|\alpha_{1o}|^{3}|\alpha_{1e}|\cos{(\phi_{1o}+\phi_{1e})}+p_{6\pm}^{2}|\alpha_{1o}|^{4}\nonumber
\\
-(p_{0\pm}(\zeta)+p_{3\pm}(\zeta)|\alpha_{1e}|^{2}+2p_{5\pm}(\zeta)|\alpha_{1o}||\alpha_{1e}|\cos{(\phi_{1o}+\phi_{1e})}
+p_{6\pm}(\zeta)|\alpha_{1o}|^{2})^{2}].\\
V_{2,3}(\zeta)=\frac{1}{<\Delta{\hat{S}_{0}}^{2}(0)>}[2q_{0}^{2}+2q_{2}^{2}+q_{3}^{2}+2q_{5}^{2}\pm
2q_{2}^{2}|\alpha_{1e}|^{4}\cos{(4\phi_{1e})}+(q_{1}^{2}+4q_{2}^{2}+q_{4}^{2})|\alpha_{1e}|^{2}+2q_{2}^{2}|\alpha_{1e}|^{4}\nonumber\\
+2(q_{1}q_{3}+2q_{2}q_{4}+2q_{4}q_{5})|\alpha_{1o}||\alpha_{1e}|\cos{(\phi_{1o}+\phi_{1e})}+4q_{2}q_{4}|\alpha_{1o}||\alpha_{1e}|^{3}\cos{(\phi_{1o}+\phi_{1e})}\pm
4q_{2}q_{4}|\alpha_{1o}||\alpha_{1e}|^{3}\cos{(\phi_{1o}-3\phi_{1e})}\nonumber\\
+4q_{2}q_{5}|\alpha_{1o}|^{2}|\alpha_{1e}|^{2}\cos{(2\phi_{1o}+2\phi_{1e})}\pm
2(q_{4}^{2}+2q_{2}q_{5})|\alpha_{1o}|^{2}|\alpha_{1e}|^{2}\cos{(2\phi_{1o}-2\phi_{1e})}\pm
4q_{4}q_{5}|\alpha_{1o}|^{3}|\alpha_{1e}|\cos{(3\phi_{1o}-\phi_{1e})}\nonumber\\
\pm
2q_{5}^{2}|\alpha_{1o}|^{4}\cos{(4\phi_{1o})}+4q_{4}q_{5}|\alpha_{1o}|^{3}|\alpha_{1e}|\cos{(\phi_{1o}+\phi_{1e})}+(q_{3}^{2}+q_{4}^{2}+4q_{5}^{2})|\alpha_{1o}|^{2}+2q_{4}^{2}|\alpha_{1o}|^{2}|\alpha_{1e}|^{2}+2q_{5}^{2}|\alpha_{1o}|^{4}\nonumber\\
-(i^{0,1}[q_{2}|\alpha_{1e}|^{2}\{e^{i2\phi_{1e}}\pm
e^{-i2\phi_{1e}}\}+q_{4}|\alpha_{1o}||\alpha_{1e}|\{e^{i\phi_{1o}-i\phi_{1e}}\pm
e^{-i\phi_{1o}+i\phi_{1e}}\}+q_{5}|\alpha_{1o}|^{2}\{e^{i2\phi_{1o}}\pm
e^{-i2\phi_{1o}} \}])^{2}].
\end{eqnarray}
Under the conditions $\phi_{1o}+\phi_{1e}=\pi$ and $k_{1}<k_{2}$
expressions (28) and (29) become more favorable for the generation
of polarization squeezed light and are evaluated for different
initial mean photon numbers in polarized coherent states. The Fig.
3-5 demonstrate the evolution of expressions (28) and (29) as
functions of the normalized interaction length $\zeta$. Fig. 3 and
Fig. 4 illustrate the relative variances of Stokes parameters
$\hat{S}_{0,2}$ and $\hat{S}_{1}$ and simulate the spontaneous
parametric down conversion process, i.e. when the photons with
frequencies $\omega_{o}$ and $\omega_{e}$ at the input of the PPNC
are each having mean single photons $|\alpha_{1o,e}|^{2}=1$ in
their orthogonal coherent states. Fig 3. shows that the variances
of the Stokes parameters $\hat{S}_{0,2}$ start with the
sub-Poissonian [3] statistics and later $\zeta>0.45$ become
super-Poissonian. The evolution of the variances of the Stokes
parameters $\hat{S}_{2}$ and $\hat{S}_{3}$ are almost the same.
So, that is why we are not demonstrating the variance of
$\hat{S}_{3}$. Moreover, the variances of the Stokes parameters
$\hat{S}_{0,2,3}$ do not differ from the same variances of the
Stokes parameters for an ordinary nonlinear crystal. The latter
one can be seen by putting the values of nonlinear coupling
coefficients $k_{1}$, $k_{2}$ equal to $0$ in the expressions (16)
and substituting $\hat{a}_{3e}(0)=0$ into (15).

The variance of the Stoke's parameter $\hat{S}_{1}$ is more
interesting due to its sub-Poissonian statistics, which is shown
in Fig. 4. It should be noted that the variance of Stoke's
parameter $\hat{S}_{1}$ for parametric down conversion process
(type II) in an ordinary nonlinear crystal is super-Poissonian.
The latter can be checked by putting the values of $k_{1}=k_{2}=0$
in the expressions (16) and substituting $\hat{a}_{3e}(0)=0$ into
(15). So, the PPNC can suppress variances of the Stokes parameters
$\hat{S}_{0,1,2}$ under certain initial values of phases
$(\phi_{1o}+\phi_{1e}=\pi)$ of polarized modes and nonlinear
coupling coefficients $(k_{1}<k_{2})$.

The expressions (28) and (29) are calculated under the same
initial conditions but with different mean photons
$|\alpha_{1o,e}|^{2}=10^{3}$ in each polarized mode. The Fig. 5
shows the evolution of variances of all the four Stokes parameters
$\hat{S}_{0,1,2,3}$. It is seen that all the variances of Stokes
parameters have sub-Poissonian statistics. After analyzing the
Fig. 3-5, it is clear that the larger the mean photon numbers in
degenerate polarized modes, the more the squeezing in the
variances of all the Stokes parameter's. Moreover, the variances
of Stokes parameters $\hat{S}_{2,3}$ are $\approx 0$, i.e. the
generation of almost Fock states.

\section{Conclusion}
In this paper we have investigated the problem of the generation
of polarization squeezed light in PPNC with second order nonlinear
susceptibility. In the case of classical and non-depleted modes at
frequencies $2\omega_{e}$ and $4\omega_{o}$, the exact solution of
the Heisenberg equations of motions is obtained. The solution is
exact quantum solution showing new possibilities for the
generation of polarized squeezed states of light. We have used
this solution to calculate the mean photon numbers of interacting
modes and variances of Stokes parameters. It is found that the
evolution of mean photon numbers of interacting modes with
frequencies $\omega_{o}$ and $\omega_{e}$ differ with the
evolution of mean photon numbers of the same interacting modes in
ordinary nonlinear crystals (degenerate parametric process (type
II)). The reasons are found, which explain the peculiarities of
degenerate parametric down conversion (type II) in PPNCs.

Also, it is shown that all the four variances of the Stokes
parameters can be squeezed (variances of the Stokes parameters are
smaller than the value of variance of coherent state)
simultaneously. Optimal initial conditions are found under which
the squeezing of variances of the Stokes parameters can be
obtained.

So, PPNCs can be good candidates for the generation of
polarization squeezed states of light, which can be used for the
realization of quantum information protocols.
\section{Acknowledgements}
One of the authors (RS) is grateful to A V Masalov for helpful
discussions.

The work was partially supported by INTAS no 01-2097.

\bigskip

\resizebox{4in}{!}{\includegraphics{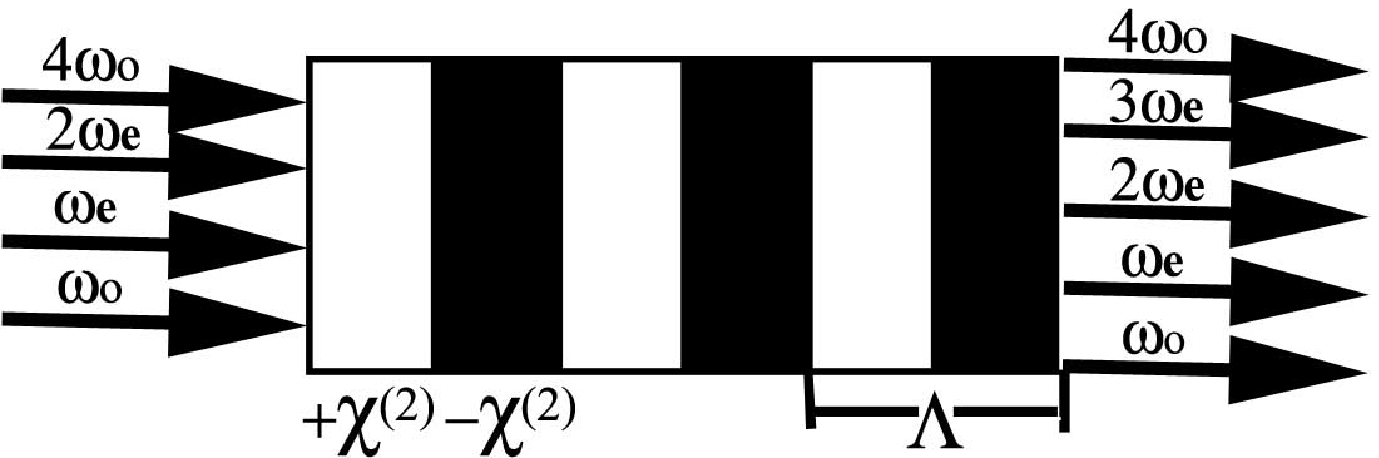}} \\Fig. 1 Sektch of
generation of polarization squeezed light in PPNC with second
order nonlinear susceptibility. The light waves involved are
described by their frequencies $\omega_{o}$, $\omega_{e}$,
$2\omega_{e}$, $3\omega_{e}$, and $4\omega_{o}$.
\bigskip

\resizebox{4in}{!}{\includegraphics{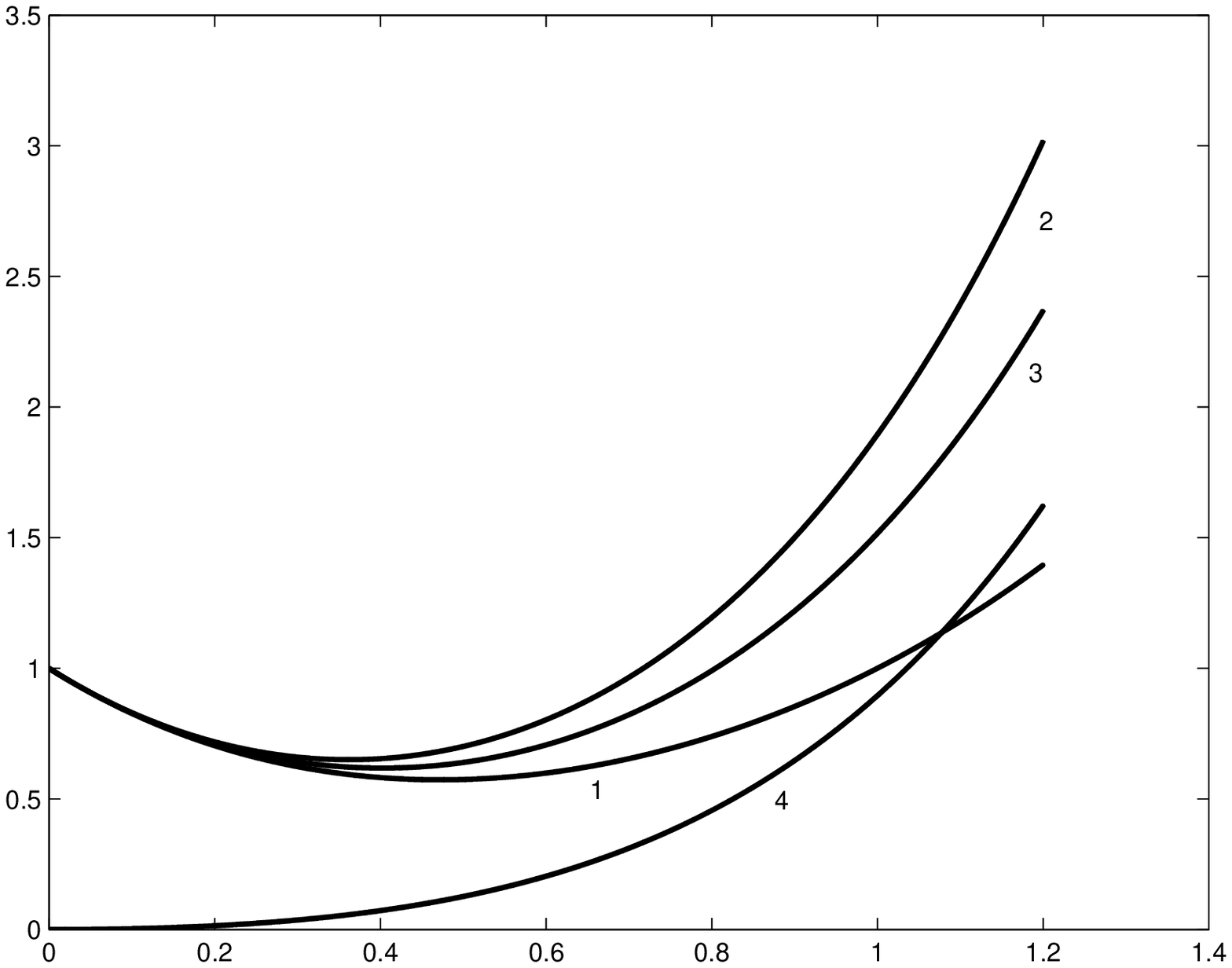}} \\Fig. 2 Mean
photon numbers $<N_{1o}(\zeta)>$ (1), $<N_{1e}(\zeta)>$ (2),
$<N_{3e}(\zeta)>$ (4) with frequencies $\omega_{o}$, $\omega_{e}$,
and $3\omega_{e}$, respectively and $<N_{1o,e}(\zeta)>
(k_{1}=k_{2}=0)$ (3), as functions of the normalized interaction
length $\zeta$. The curves (1-4) are calculated corresponding to
the initial photon numbers $<N_{1o,e}(0)>=1$ and $<N_{3e}(0)>=0$
\bigskip

\resizebox{4in}{!}{\includegraphics{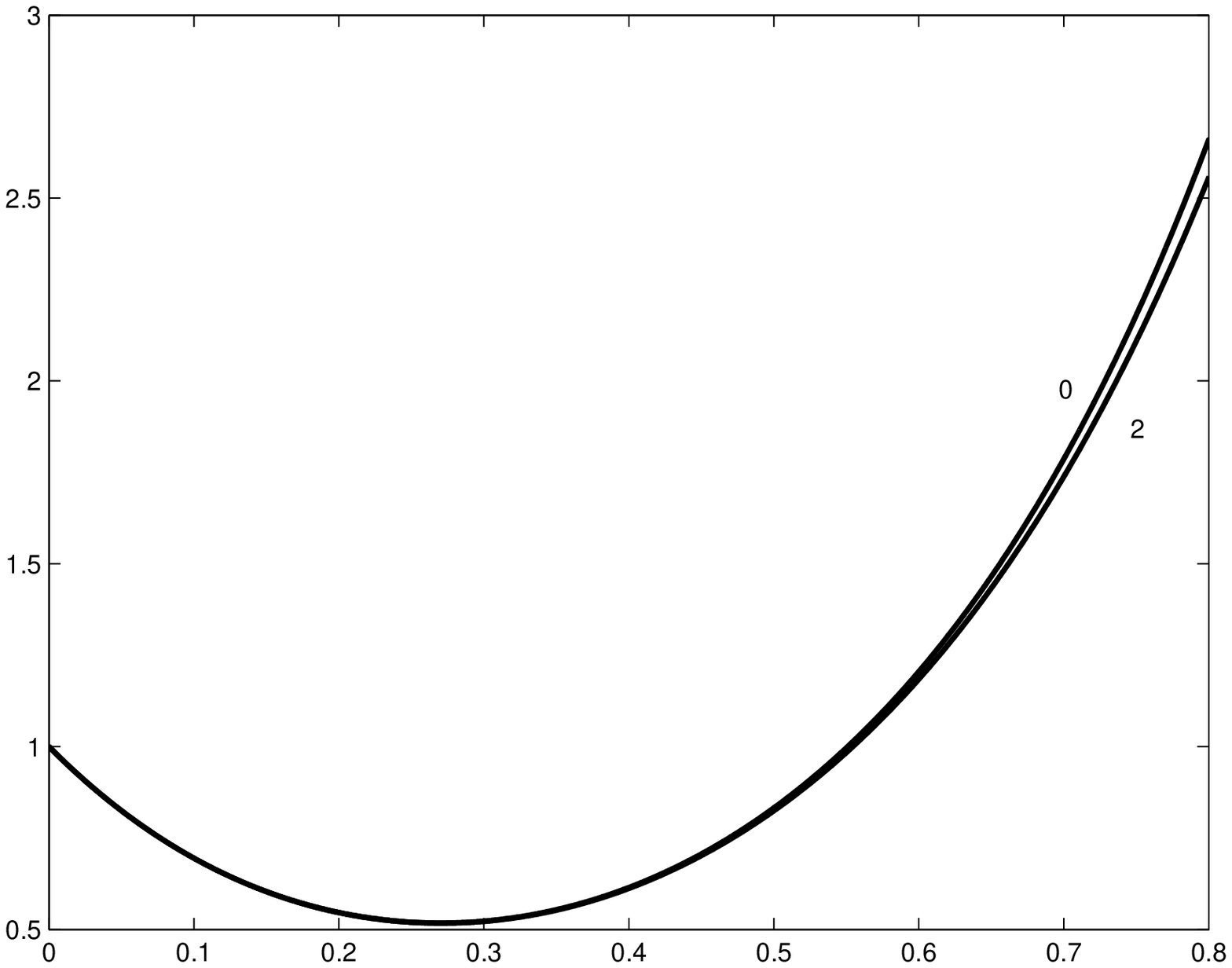}} \\Fig. 3 Normalized
variances $V_{0}(\zeta)$ (0) and $V_{2}(\zeta)$ (2) of the Stokes
parameters $\hat{S}_{0}(\zeta)$ and $\hat{S}_{2}(\zeta)$. Curves
(0 and 2) are calculated corresponding to the initial photon
numbers $<N_{1o,e}(0)>=1$ and $<N_{3e}(0)>=0$
\bigskip

\resizebox{4in}{!}{\includegraphics{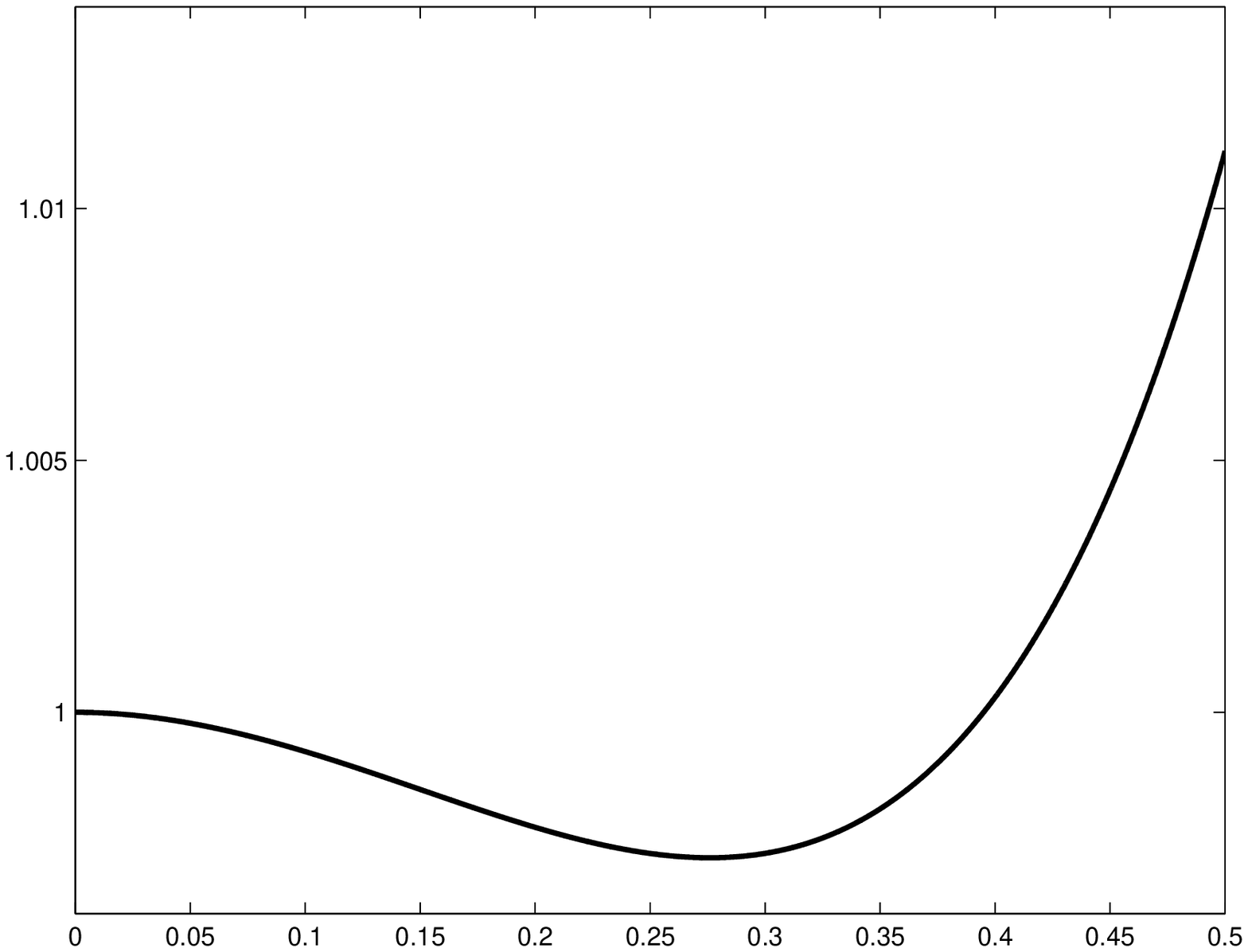}} \\Fig. 4 Normalized
variance $V_{1}(\zeta)$ of the Stoke's parameter
$\hat{S}_{1}(\zeta)$. Curve is calculated corresponding to the
initial photon number $<N_{1o,e}(0)>=1$ and $<N_{3e}(0)>=0$
\bigskip

\resizebox{4in}{!}{\includegraphics{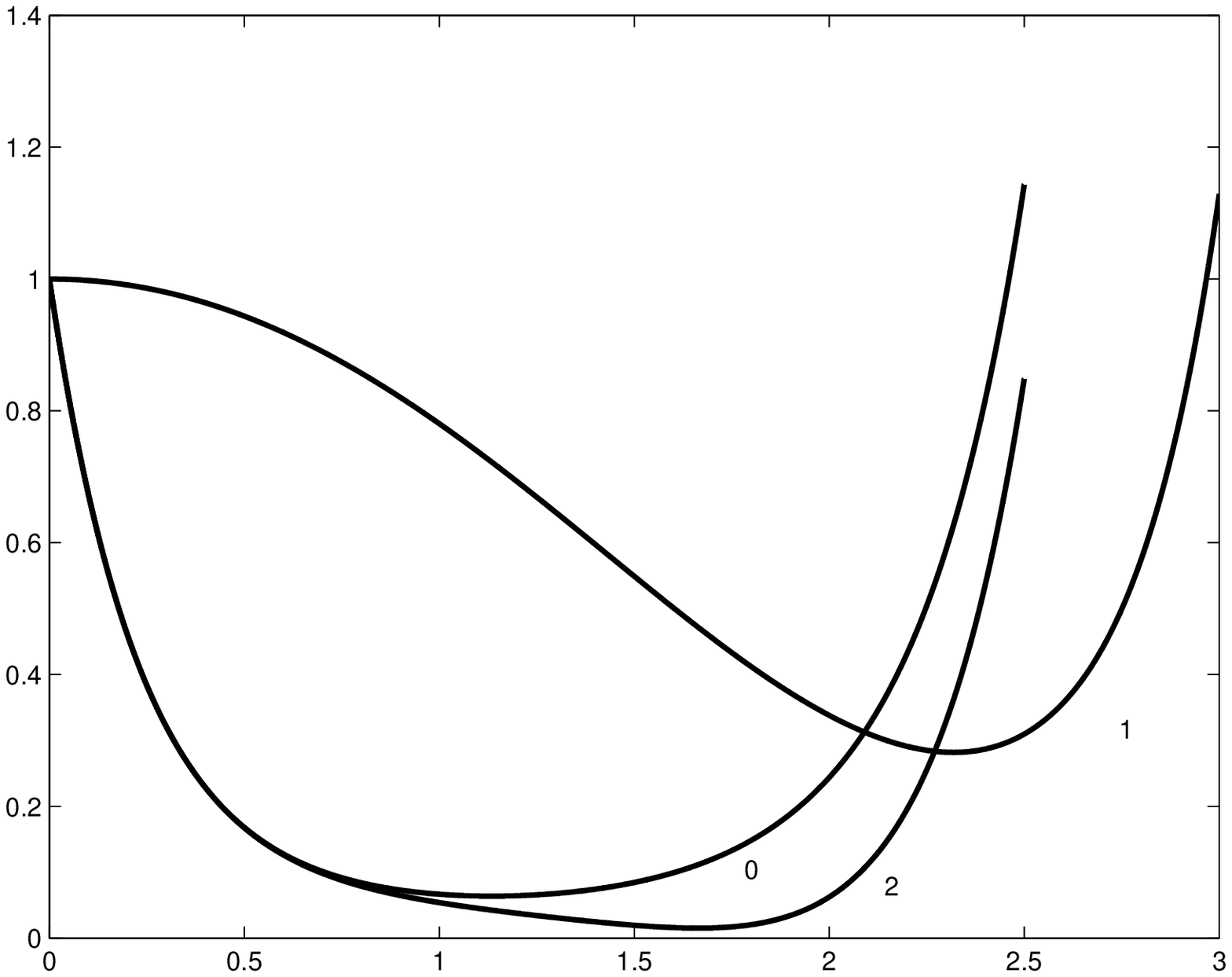}} \\Fig. 5 Normalized
variances $V_{0}(\zeta)$ (0) and $V_{1}(\zeta)$ (1), and
$V_{2}(\zeta)$ (2), of the Stokes parameters $\hat{S}_{0}(\zeta)$,
$\hat{S}_{1}(\zeta)$ and $\hat{S}_{2}(\zeta)$. Curves (0-2) are
calculated corresponding to the initial photon numbers
$<N_{1o,e}(0)>=10^3$ and $<N_{3e}(0)>=0$

\end{document}